\documentclass[fleqn,usenatbib]{mnras}

\usepackage{newtxtext,newtxmath}
\usepackage[T1]{fontenc}

\usepackage{graphicx}	% Including figure files
\usepackage{amsmath}	% Advanced maths commands
\title[Cusp formation from peak collapse]{Prompt cusp formation from the gravitational collapse of peaks in the initial cosmological density field}

\author[Simon D.M. White]{
Simon D.M. White,$^{1}$\thanks{E-mail: swhite@mpa-garching.mpg.de }\\
$^{1}$Max Planck Institute for Astrophysics, Karl-Schwarzschild-Strasse 1, 85740 Garching, Germany\\
}

\date{Accepted XXX. Received YYY; in original form ZZZ}

\pubyear{2022}

\begin{document}
\label{firstpage}
\pagerange{\pageref{firstpage}--\pageref{lastpage}}
\maketitle

\begin{abstract}
I present an analytic model for the early post-collapse evolution of a spherical density peak on the coherence scale of the initial fluctuations in a universe filled with collisionless and pressure-free ``dust''. On a time-scale which is short compared to the peak's collapse time $t_0$, its inner regions settle into an equilibrium cusp with a power-law density profile, $\rho\propto r^{-12/7}$. Within this cusp, the circular orbit period $P$ at each radius is related to the enclosed mass $M$ by $P = t_0 (M/M_c)^{2/3}$ where $M_c$ is a suitably defined characteristic mass for the initial peak. The relaxation mechanism which produces this cusp gives insight into those which are active in high-resolution simulations of first halo formation in Cold or Warm Dark Matter universes, and, indeed, a simple argument suggests that the same power-law index $\gamma=-12/7$ should describe the prompt cusps formed during the collapse of generic peaks, independent of any symmetry assumption. Further work is needed to investigate the additional factors required to explain the slightly flatter exponent, $\gamma\approx -1.5$, found in high-resolution numerical simulations of peak collapse.
\end{abstract}


%%%%%%%%%%%%%%%%%%%%%%%%%%%%%%%%%%%%%%%%%%%%%%%%%%

%%%%%%%%%%%%%%%%% BODY OF PAPER %%%%%%%%%%%%%%%%%%

\section{Introduction}

Recent high-resolution simulations of the collapse and evolution of the first generation of dark matter haloes have found that shortly after collapse their inner structure is well fit by a power-law density profile $\rho(r) = A r^{\gamma}$ with $\gamma\approx -1.5$ \citep{Ishiyama2010,Anderhalden2013,Ishiyama2014,Angulo2017,Ogiya2018,Delos2018b,Delos2018a,Colombi2021,Delos2022}. In particular, \cite{Delos2022} show that the constant $A$ can be estimated to within about 10\% for a given halo from the characteristic size and the collapse time of the initial density peak associated with the halo. In both
Cold and Warm Dark Matter cosmologies the mass and size of the first haloes grow very rapidly after initial collapse, and their density profiles evolve to take on the familiar non-power-law forms typical of more massive haloes at later times \citep{Navarro1996,Navarro2004}.
The highest resolution dark matter-only simulations of galaxy and cluster haloes find  effective profile slopes significantly shallower than $\gamma=-1.5$ at the smallest resolved radii \citep[e.g.][]{Springel2008, Gao2012}. It is nevertheless plausible that at even smaller radii the profiles steepen again because the initial cusp is preserved \citep[see, for example][]{Delos2022}. Despite the substantial body of work referenced above, there is still no explanation for {\it why} the collapse of a density peak should produce a cusp with $\gamma\approx -1.5$ on a timescale much shorter than the peak's initial collapse time. The goal of this paper is to provide such an explanation.

The first analytic model for halo formation was that of \cite{Gunn1972} who imagined imposing an additional point mass on an otherwise unperturbed Einstein-de Sitter universe filled with pressure-free collisionless ``dust''. They showed that surrounding material would fall back onto the perturber in such a way as to establish a dark matter halo with $\rho\propto r^{\gamma}$ where $\gamma=-9/4$. More general similarity solutions for spherical collapse were published by \cite{Fillmore1984ApJ}, although they found no solutions with $\gamma>-2$ because they assumed purely radial orbits.  As shown by \cite{White1996}, allowing orbits with non-zero (but randomly oriented) angular momentum leads to similarity solutions with halo profile slopes in the full range  $0<\gamma<-3$. All these solutions have the property that the circular orbit period at the radius enclosing mass $M$ in the equilibrium halo is proportional to the infall time onto the halo of the shell which enclosed
mass $M$ in the early universe. A similar relationship between infall history and halo structure underlies the model for halo profile formation proposed by \cite{Ludlow2013}. Such ideas cannot explain cusp formation immediately after the initial collapse of a density peak, since all the mass shells contributing to the cusp fall onto the halo at very nearly the same time. As \cite{Delos2022} note, a different idea is needed.

\section{The collapse of a spherical peak}
\label{sec:spherical}

Consider a spherical density peak in the linear overdensity field of an otherwise unperturbed Einstein-de Sitter universe filled with pressure-free, collisionless ``dust''. At early times $t \ll t_0$ the density in the neighborhood of the peak can be written to lowest order as
\begin{equation}
    \rho(q,t) = \bar{\rho}(t)\big[1 + \delta(q,t)\big] = \bar{\rho}(t)\big[1 +\delta_c(t)(1 - q^2/6R^2)\big],
    \label{eq:linpeak}
\end{equation}
where $q=r/a$ is a Lagrangian radial coordinate, $\bar{\rho}= 1/(6\pi Gt^2) = \bar{\rho}_0(t/t_0)^{-2} = \bar{\rho}_0a^{-3}$ is the mean cosmological background density, $a=(t/t_0)^{2/3}$ is the cosmological expansion factor, $\delta_c = 1.686a$ is the central linear overdensity of the peak, $R = |\delta/\nabla^2\delta|^{1/2}$ (also evaluated at peak centre) is a characteristic Lagrangian size for the peak, and the subscript 0 indicates quantities evaluated when the centre of the peak collapses to (nominally) infinite density. At these early times the mass within $q$ is $M(q) = 4\pi\bar{\rho}_0 q^3/3 = M_0(q/R)^3$ where $M_0=4\pi\bar{\rho}_0 R^3/3$. The mean enclosed overdensity within radius $q$ can thus be written
\begin{equation}
    \bar\delta(q,t) = 1.686a(1 - q^2/10R^2) = 1.686a(1 - 0.1(M/M_0)^{2/3}).
    \label{eq:encloverden}
\end{equation}
During the initial expansion and collapse phases, the radial ordering of mass shells is preserved until just before each shell collapses to the origin. As a result, collapse occurs very nearly at the time when the enclosed linear overdensity of the shell extrapolates to 1.686. Thus, to lowest order in $M/M_0$, the collapse time of the shell is
\begin{equation}
    t_c(M) = t_0(1 + 0.15(M/M_0)^{2/3}). 
    \label{eq:colltime}
\end{equation}
At times shortly before $t_c(M)$ the infall velocity of the shell satisfies $0.5(dr/dt)^2 = GM/r$, so that
\begin{equation}
    r^{3/2} = 3\sqrt{GM/2}~(t_c(M) - t)
    \label{eq:lateinfall}
\end{equation}
or
\begin{equation}
    r^{3/2} = 3\sqrt{GM/2}~t_0(0.15(M/M_0)^{2/3} - \Delta t/t_0),
\end{equation}
where $\Delta t = t - t_0$. This expression then simplifies to
\begin{equation}
    (r/R)^{3/2} = (M/M_0)^{1/2}(0.15(M/M_0)^{2/3} - \Delta t/t_0)
    \label{eq:Mrt-rel}
\end{equation}
with approximate validity for $0<\Delta t/t_0<0.15(M/M_0)^{2/3}$ and $M< M_0$, thus for a relatively short period after the initial collapse of the peak. The approximation arises because just before reaching the origin, an infalling shell will cross previously collapsed shells; the enclosed mass will then drop below $M$.

It is convenient to work with scaled variables and to set the origin of time to $t_0$. Defining $r^\prime = r/R$, $m = M/M_0$ and $s=\Delta t/t_0$, equation \ref{eq:Mrt-rel} becomes
\begin{equation}
    {r^\prime}^{3/2} = m^{1/2}(0.15m^{2/3} - s),
    \label{eq:Mrt-relsc}
\end{equation}
with $0<s<0.15m^{2/3}$ and $m<1$.
At given time $s$, this equation can be viewed as defining the relation between radius $r^\prime$ and the enclosed mass $m(r^\prime,s)$.  As $r^\prime$ tends to zero, $m$ tends to $m_c(s) = (s/0.15)^{3/2}$. This is just the inverse of equation \ref{eq:colltime}, which defines the collapse time of the shell which initially enclosed mass $m$ as $s_c(m)= 0.15 m^{2/3}$.

Now consider a particular mass shell which initially contained mass $M^\prime$ and so first collapsed at time $t = t_0 + 0.15 (M^\prime/M_0)^{2/3}$. As this shell re-expands after collapse, its equation of motion is
\begin{equation}
    \frac{d^2r}{dt^2} = - \frac{GM(r,t)}{r^2} 
\end{equation}
which in our scaled variables becomes
\begin{equation}
    \frac{d^2r^\prime}{ds^2} = - \frac{2m(r^\prime,s)}{9{r^\prime}^2}, 
    \label{eq:EoM}
\end{equation}
During the last phase of its initial collapse, the radius of the shell satisfies equation \ref{eq:lateinfall}, so it is natural to assume that the immediate post-collapse motion is the time-reverse of this which may be written in our scaled variables as
\begin{equation}
    {r^\prime}^{3/2} = {m^\prime}^{1/2}\big(s - s_c(m^\prime)\big)
    \label{eq:EoM_ic},
\end{equation}
where $m^\prime = M^\prime/M_0$. This provides initial conditions for integrating equation \ref{eq:EoM} to obtain the subsequent outward
trajectory. Notice that for $s>s_c(m^\prime)$, equation \ref{eq:Mrt-relsc} implies $m>m^\prime$ for all $r^\prime$. Thus the deceleration of the shell on its outward trajectory is stronger than the acceleration it experienced at the same radius during infall. As a result, the outward motion will stop and the shell will turn around for the second time at a smaller radius than its first apocentre. It is this asymmetry between infall and re-expansion which gives rise to the final cusp in the halo density profile.

In the regime we are considering, the trajectories followed by different mass shells are self-similar. Let us change variables again to $s^{\prime\prime} = s/{m^\prime}^{2/3}$ and $r^{\prime\prime} = r^\prime/{ m^\prime}^{7/9}$. In addition, let us define $f = m(r^\prime,s)/m^\prime$. Equation \ref{eq:Mrt-relsc} then takes the form
\begin{equation}
{r^{\prime\prime}}^{3/2} = f^{1/2}(0.15f^{2/3} - s^{\prime\prime}),
\end{equation}
which determines $f(r^{\prime\prime}, s^{\prime\prime})$. This can be inserted into the equation of motion \ref{eq:EoM} to give
\begin{equation}
   \frac{d^2r^{\prime\prime}}{d{s^{\prime\prime}}^2} = - \frac{2f(r^{\prime\prime},s^{\prime\prime})}{9{r^{\prime\prime}}^2}.
    \label{eq:EoM_fin}  
\end{equation}
 The initial solution at small radius (equation \ref{eq:EoM_ic}) becomes
\begin{equation}
    r^{\prime\prime} = (s^{\prime\prime}- 0.15)^{2/3}.
    \label{eq:EoM_icfin}
\end{equation}
In these variables the post-collapse trajectory is thus independent of $m^\prime$.
Integrating these equations numerically shows the second apocentre to occur when $s^{\prime\prime} = 0.199$, $r^{\prime\prime} = 0.087$ and $f=1.75$. At this time, the mass enclosed by a shell is thus about 1.75 times the (constant) mass enclosed during its initial expansion and contraction. In the original variables, the time and radius of second apocentre are 
\begin{equation}
\begin{aligned}
    &t_{\rm max}(M^\prime) = t_0\big(1 +0.199(M^\prime/M_0)^{2/3}\big),\\ &r_{\rm max}(M^\prime) = 0.087R~(M^\prime/M_0)^{7/9}.\\
    \end{aligned}
    \label{eq:scaling}
\end{equation}
Since $r_{\rm max}$ increases relatively rapidly with $M^\prime$, the equilibrium post-collapse structure is established from inside out and on a timescale which is much shorter than $t_0$. As a result, the gravitational force at radii $r<r_{\rm max}$  is expected to evolve little at times $t>t_{\rm max}$, and the enclosed mass within radius $r$ in the resulting object will scale as $M(r) \propto r^{9/7}$, so that $\rho \propto r^{-12/7}$ is its equilibrium density profile.\footnote{At the instant of first collapse, $s=0$, Equation~\ref{eq:Mrt-relsc} implies $m\propto {r^{\prime}}^{9/7}$, hence $\rho\propto r^{-12/7}$. This was noticed, apparently independently, by \cite{Penston69}, \cite{Nakamura85} and \cite{Gurevich95}. Although this density exponent is the same as found here, the binding energy distribution of the infalling shells is quite different at this instant from their late-time equilibrium distribution since violent relaxation has yet to occur.}  The proportionality constant in this relation can be estimated by noting that the time between first collapse and second apocentre for the shell initially enclosing mass $M$ is $t_{\rm max} - t_c = 0.049(M/M_0)^{2/3}t_0$. This should be roughly half the period of the circular orbit in the equilibrium object enclosing mass $M$.\footnote{This is justified by noting that in standard spherical infall models, half the circular orbit period at $r_{200c}$, the conventional halo boundary, is 0.94 of the time since first apocentre for the shell currently accreting onto the halo.} Adopting this assumption implies a density profile
\begin{equation}
  \rho(r) = 20.6\bar{\rho}_0(r/R)^{-12/7}.
  \label{eq:cusp}
\end{equation}
This cusp is expected to extend out to $M\sim M_0$, thus to a limiting radius $r_{\rm lim} \sim 0.049R$.
For comparison, \cite{Delos2022} found 
\begin{equation}
  \rho(r) \approx 24\bar{\rho}_0(r/R)^{-1.5}
\end{equation}
for the central cusps which formed in their 12 high-resolution simulations of peak collapse from more realistic (and generically non-spherical) cosmological initial conditions. For $r=r_{\rm lim}$ this implies a density about 40\% smaller than that given by equation \ref{eq:cusp}. This appears an acceptable level of agreement given the uncertainty in the assumptions used to estimate the coefficient in this equation.
\section{Discussion}
\label{sec:discussion}

There are several reasons to be sceptical of the results of  the last section. For a perfectly spherical system, all motions are purely radial, as assumed in the
analysis leading to the relations \ref{eq:scaling}. An equilibrium cusp  with purely radial orbits must, however, have $\gamma$ less than or similar to $-2$ regardless of the distribution of particle binding energies, because of the requirement that all orbits pass through $r=0$ \citep[see, for example, the similarity solutions of][]{Fillmore1984ApJ}\footnote{Note that when spherical symmetry is unbroken, the similarity solution found here is identical to the case $\epsilon = 4/9$ of \cite{Fillmore1984ApJ} except that the time variable is not cosmic time but rather time since the instant of peak collapse.}. In this case, the scaling argument following equation \ref{eq:scaling} fails unless particles acquire some angular momentum before settling onto their final orbits. One possible mechanism to generate non-radial motions is the radial orbit instability, which would operate during the first few orbits after collapse
\citep[e.g.][]{Barnes1986}. The tidal field at density peaks is generically anisotropic, and so all particle orbits will acquire a non-radial component during their initial collapse unless they lie along one of the principal axes of the tidal tensor. This generates a seed for the radial orbit instability, but may also cause large enough tangential motions for the instability to be suppressed. In either case, the prediction of $\gamma =-12/7$ for the cusp would be restored \citep[as in the model proposed by] [to solve the analogous problem in earlier similarity solutions]{White1996}. 

Another problem with the analysis of the last section is that at second apocentre the binding energy of every mass shell is larger than its value at infall and is assumed to be conserved thereafter. Gravitational dynamics is Hamiltonian, however, so an appropriately defined total energy should be conserved. Part of this problem comes from the fact that the analysis includes the additional enclosed mass which a shell sees as it moves from collapse to second apocentre, but neglects the corresponding drop in enclosed mass that it must experience in the last stages of its initial collapse. In addition, the mass distribution within $r_{\rm max}$  must change significantly during the half orbit after second apocentre as the shell crosses others with slightly later initial infall times which are on their way out to larger second apocentres. A proper treatment of these effects would not change the scaling properties found above, but would lead to different coefficients in  equations \ref{eq:scaling}.  

The most obvious reason to doubt the applicability of the arguments of Section \ref{sec:spherical} is, however, the assumption of spherical symmetry. As demonstrated graphically and in considerable detail by \cite{Delos2022}, the morphology of the initial collapse of typical cosmological density peaks is varied and complex; material passes through a network of caustics and higher order singularities which is very far from spherical.  However, these authors also  show several cases where the material that ends up in different, nested radial bins in the final cusp comes from  a similarly nested set of roughly ellipsoidal shells in the (Lagrangian) initial density field. This is in exact analogy with the spherical case, motivating the following, general argument which suggests that generic, non-spherical peak collapse may also lead to prompt central cusps satisfying equation \ref{eq:cusp}.

Imagine labelling each mass element in the initial (Lagrangian) cosmological density field by a collapse time $t_c({\bf q})$, defined as the first time when the element lies within a fully collapsed object (e.g. an object which has collapsed along all three principal axes). Viewed as a function of ${\bf q}$, the field $t_c({\bf q})$ is expected to have a similar coherence length to the linear overdensity field and to have minima close to density peaks. Consider a minimum at ${\bf q}_0$ with depth $t_0$. In its vicinity, one can  expand $t_c({\bf q})$ to second order in $\Delta{\bf q} = {\bf q} - {\bf q}_0$. Surfaces of constant $\Delta t = t_c - t_0$ are then similar ellipsoids centred on ${\bf q}_0$, implying that $\Delta t$ and the enclosed mass $M$ scale as the square and the cube of ellipsoid size, respectively. Hence $\Delta t\propto M^{2/3}$ just as in equation \ref{eq:colltime}. At time $t_0 + \Delta t$ the central collapsed object has mass $M\propto(\Delta t)^{3/2}$, all of which has assembled since $t_0$. If gravitational interactions cause this material to settle onto equilibrium orbits with periods proportional to $\Delta t$, then a cusp will form in which circular orbit period and enclosed mass satisfy $P\propto M_{\rm enc}^{2/3}$.  This implies $\rho\propto r^{-12/7}$, just as in the spherical model. 

It seems, therefore, that prompt cusps form during the collapse of a smooth density peak because material from the inner regions settles onto orbits with periods which are proportional not to its collapse time, as in the standard similarity solutions for cosmological halo formation, but rather to the {\it difference} between its collapse time and that of the very centre of the perturbation. It remains to be seen whether this idea captures all the salient features of violent relaxation in the inner regions of peak collapse, or whether additional factors, for example an extended period of violent relaxation and mixing in the innermost regions, are needed to explain the difference between the power-law exponent $\gamma\approx -1.5$ found in high-resolution simulations and the exponent $\gamma=-12/7 = 1.71$ given by the arguments of this paper.

%%%%%%%%%%%%%%%%%%%%%%%%%%%%%%%%%%%%%%%%%%%%%%%%%%

\section*{Acknowledgements}

I thank Sten Delos for convincing me of the reality of prompt cusps as well as of the need to explain why and how they form and why their density exponents appear universal. 

%%%%%%%%%%%%%%%%%%%%%%%%%%%%%%%%%%%%%%%%%%%%%%%%%%
\section*{Data Availability}
 
No new data were generated or analysed in support of this research.



\bibliographystyle{mnras}
\bibliography{cusp} % if your bibtex file is called example.bib

%\begin{thebibliography}{99}
%\end{thebibliography}

%%%%%%%%%%%%%%%%%%%%%%%%%%%%%%%%%%%%%%%%%%%%%%%%%%

%%%%%%%%%%%%%%%%% APPENDICES %%%%%%%%%%%%%%%%%%%%%
\appendix
% Don't change these lines
\bsp	% typesetting comment
\label{lastpage}
\end{document}